\documentclass[showpacs,preprintnumbers,amsmath,amssymb,floatfix,elsart]{revtex4}

\usepackage{graphicx}
\usepackage{dcolumn}
\usepackage{bm}

\newcommand{\beq}{\begin{equation}}
\newcommand{\eeq}{\end{equation}}
\newcommand{\bey}{\begin{eqnarray}}
\newcommand{\eey}{\end{eqnarray}}

\begin{document}

\title{On the stability of thin-shell wormholes}

\author{Peter K.F. Kuhfittig}
\email{kuhfitti@msoe.edu}\affiliation{ Department of
Mathematics, Milwaukee School of Engineering, Milwaukee, Wisconsin
53202-3109, USA}


\begin{abstract}\noindent
A thin-shell wormhole is theoretically constructible
by surgically grafting together two Schwarzschild
spacetimes using the so-called cut-and-paste technique.
By describing such a wormhole as the limiting case
of a spherical shell, it is shown that the
structure must be unstable to linearized radial
perturbations.  Some earlier studies by the author
et al. have shown, however, that under certain
conditions, thin-shell wormholes can be stable.

\end{abstract}


\maketitle

\section{Introduction}\label{S:Introduction}\noindent
It is well known that the thin-shell wormholes due to
Visser \cite{PV95} are constructed by the so-called
cut-and-paste technique: start with two copies of
Schwarzschild spacetime and remove from each manifold
the four-dimensional regions
\begin{equation}
   \Omega^{\pm}= \{r\le a\,|\,a>2M\},
\end{equation}
where $a$ is a constant.  We now identify (in the
sense of topology) the timelike hypersurfaces
\begin{equation}
   \partial\Omega^{\pm}= \{r=a\,|\,a>2M\}.
\end{equation}
The resulting manifold is geodesically complete and
has two asymptotically flat regions connected by
the throat $r=a$.

The subsequent stability analysis depends on the
assumption that the radius of the throat is a
function of time, so that the induced metric is
\begin{equation}
   ds^2=-d\tau^2+[a(\tau)]^2( d\theta^2+
   \text{sin}^2\theta\, d\phi^2),
\end{equation}
where $\tau$ is the proper time on the junction surface.
So it becomes convenient to denote $da/d\tau$ by
$\dot{a}$.

The stability analysis is carried out by plotting the
radius $r=a$ against $\beta^2$, where $\beta$ is
usually interpreted as the speed of sound.  There
are regions of stability for $\beta^2\ge 3/2+\sqrt{5}$ and
$\beta^2\le -1/2$.  While $\beta$ would normally be
confined to the interval $(0,1]$, Visser
notes that we are dealing with exotic matter whose
properties are not well understood.  In fact,
$\beta^2$ could be just a convenient parameter.

The purpose of this note is to study a wormhole
restricted to a spherical shell that can be made
arbitrarily thin and having the appropriate
surface stresses, thereby approximating a thin
shell due to Schwarzschild surgery, and to
show that such a wormhole is indeed unstable
to linearized radial perturbations.
Connections to other studies of thin-shell
wormholes are discussed at the end.  (A more
general discussion of the properties of
thin-shell wormholes and their stability
can be found in Refs. \cite{LL08} and
\cite{DL10}.)

\section{Wormhole structure}\noindent
We start with the static and spherically
symmetric line element
\begin{equation}\label{E:line1}
   ds^2=- e^{2\Phi(r)} dt^2+\frac{dr^2}{1-b(r)/r}
   +r^2( d\theta^2+\text{sin}^2\theta\, d\phi^2).
\end{equation}
(We are using geometrized units: $c=G=1.$)  Here
$\Phi(r)$ is called the \emph{redshift function}, which
must be everywhere finite to prevent an event horizon.
The function $b(r)$ is called the \emph{shape function}
because it determines the shape of the wormhole when
viewed, for example, in an embedding diagram.  The
shape function must satisfy the following conditions:
(1) $b(r_0)=r_0$, where $r=r_0$ is the \emph{throat}
of the wormhole, (2) the \emph{flare-out} condition
$b'(r_0)<1$, and (3) $b(r)<r$ for $r>r_0$.  The
flare-out condition can only be satisfied by violating
the null energy condition \cite{MT88}.

The Einstein field equations are stated next:
\begin{equation}\label{E:Einstein1}
  \frac{b'}{r^2}= 8\pi \rho,
\end{equation}
\begin{equation}\label{E:Einstein2}
   -\frac{b}{r^3}+2\left(1-\frac{b}{r}\right)
   \frac{\Phi'}{r}=
8\pi p_r,
\end{equation}
and
\begin{equation}\label{E:Einstein3}
   \left(1-\frac{b}{r}\right)\left[\Phi''+(\Phi')^2
   -\frac{b'r-b}{2r(r-b)}\Phi'-\frac{b'r-b}{2r^2(r-b)}
   +\frac{\Phi'}{r}\right]\\ =8\pi p_t.
\end{equation}
For now we will assume that the density $\rho$ is constant
and that the wormhole material is confined to the
spherical shell $r_0\le r\le a$,
i.e.,
\begin{equation}\label{E:rho}
   \rho(r)=
   \begin{cases}
    \rho_0,&r_0\le r\le a\\
    \,\,\,0, &\,r>a,
    \end{cases}
\end{equation}
where $\rho_0$ is a constant and $r=r_0$ is the throat.
(This form of $\rho(r)$ was also considered by Sushkov
\cite{sS05} in the discussion of wormholes supported by
phantom energy.)  Now from Eq. (\ref{E:Einstein1}), we
obtain the shape function
\begin{equation}\label{E:shape2}
  b(r)=\frac{8\pi r^3}{3}\rho_0-
  \frac{8\pi r_0^3}{3}+r_0.
\end{equation}
Observe that $b(r_0)=r_0$.  Since we are using
geometrized units, we can expect $\rho_0$ to be quite
small compared to, say, $r_0$.  Hence
\begin{equation}\label{E:flare}
   b'(r_0)=8\pi r_0^2\,\rho_0<1,
\end{equation}
thereby satisfying the flare-out condition.

\section{Junction to an external vacuum solution}\noindent
Since the wormhole material is cut off at $r=a$, the
interior solution must be matched to the exterior
Schwarzschild solution
\begin{equation}\label{E:line3}
   ds^2=-\left(1-\frac{2M}{r}\right)dt^2
   +\frac{dr^2}{1-2M/r}\\+r^2( d\theta^2
   +\text{sin}^2\theta\, d\phi^2).
\end{equation}
Since the metric coefficients are continuous, we
must have in view of Eq. (\ref{E:line1}),
\begin{equation}\label{E:mass}
   M=\frac{1}{2}b(a)=\frac{4\pi a^3}{3}\rho_0
   -\frac{4\pi r_0^3}{3}\rho_0+\frac{1}{2}r_0
\end{equation}
and
\begin{equation}
   \Phi(a)=\frac{1}{2}\text{ln}
   \left(1-\frac{2M}{a}\right).
\end{equation}
(The components $g_{\theta\theta}$ and
$g_{\phi\phi}$ are already continuous
due to the spherical symmetry.)

At this point we need to turn our attention to
surface stresses, since these will place some
severe restrictions on the wormhole geometry.  To
see why, let us recall the Lanczos equations
\cite{fL04}
\begin{equation}\label{E:sigma1}
     \sigma=-\frac{1}{4\pi}\kappa^{\theta}_
     {\phantom{\theta}\theta}
\end{equation}
and
\begin{equation}
   \mathcal{P}=\frac{1}{8\pi}(\kappa^{\tau}
   _{\phantom{\tau}\tau}+\kappa^{\theta}
   _{\phantom{\theta}\theta}),
\end{equation}
where $\kappa_{ij}=K^{+}_{ij}-K^{-}_{ij}$ and
$K_{ij}$ is the extrinsic curvature.  According to
Ref. \cite{fL04},
\begin{equation*}
   \kappa^{\theta}_{\phantom{\theta}\theta}=\frac{1}{a}
   \sqrt{1-\frac{2M}{a}}-\frac{1}{a}
      \sqrt{1-\frac{b(a)}{a}}.
\end{equation*}
So by Eq. (\ref{E:sigma1}),
\begin{equation}\label{sigma2}
   \sigma=-\frac{1}{4\pi a}
   \left(\sqrt{1-\frac{2M}{a}}-
      \sqrt{1-\frac{b(a)}{a}}\right).
\end{equation}
In view of Eq. (\ref{E:mass}) one could reasonably
expect that $\sigma=0$.  However, part of the
junction formalism is to assume that the junction
surface $r=a$ is an infinitely thin surface having
a nonzero density that may be positive or negative.
Hence its mass is given by
\begin{equation}\label{E:ms}
  m_s=4\pi a^2\sigma=-a\left(\sqrt{1-\frac{2M}{a}}
  -\sqrt{1-\frac{b(a)}{a}}\right).
\end{equation}
As a check, if $\sigma(a)<0$ (resp. $\sigma(a)>0$),
then $M<\frac{1}{2}b(a)$ (resp. $M>\frac{1}{2}b(a)$), 
and $m_s$ is negative (resp. positive).

Next,
\[
   K^{\tau\,+}_{\phantom{\tau}\tau}=\frac{M/a^2}
   {\sqrt{1-2M/a}}
\]
and
\[
   K^{\tau\,-}_{\phantom{\tau}\tau}=
   \Phi'(a)\sqrt{1-\frac{b(a)}{a}}.
\]
So the surface pressure is given by
\begin{multline}\label{E:pressure1}
   \mathcal{P}=\frac{1}{8\pi}\left[\frac{M/a^2}
   {\sqrt{1-2M/a}}-\Phi'(a)
   \sqrt{1-\frac{b(a)}{a}}
   +\frac{1}{a}\sqrt{1-\frac{2M}{a}}
   -\frac{1}{a}\sqrt{1-\frac{b(a)}{a}}\right ]\\
   =\frac{1}{8\pi}\left [\frac{M/a^2-\Phi'(a)
   \sqrt{1-2M/a}\sqrt{1-b(a)/a}}
   {\sqrt{1-2M/a}}+\frac{1}{a}\sqrt{1-\frac{2M}{a}}
   -\frac{1}{a}\sqrt{1-\frac{b(a)}{a}}\right ]\\
   \approx \frac{1}{8\pi}
   \frac{M/a^2-\Phi'(a)(1-2M/a)}{\sqrt{1-2M/a}}
\end{multline}
since $b(a)\approx 2M$.  Before substituting
the expression for $M$ from Eq. (\ref{E:mass}),
recall that $\rho_0$ is relatively small, so that
\begin{equation}\label{E:approxP}
   \mathcal{P}\approx\frac{1}{8\pi\sqrt{1-2M/a}}
   \left(\frac{r_0/2}{a^2}-\Phi'(a)+\Phi'(a)
   \frac{r_0}{a}\right).
\end{equation}
It is easily shown that $\mathcal{P}>0$ if,
and only if, $\sigma(a)<0$.  According to Eq.
(\ref{E:approxP}), $\mathcal{P}$ remains positive
even if $a$ is arbitrarily close to $r_0$, i.e.,
\[
    \text{lim}_{a\rightarrow r_0}\mathcal{P}>0.
\]
It follows that $\sigma(a)<0$.

To draw a more general conclusion, suppose we
assume that $a$ is indeed close to $r_0$, thereby
forming a thin spherical shell.  On the small
interval $[r_0,a]$, any function $\rho(r)$ is
approximately constant, so that the assumption
of a constant density can be omitted.  Moreover,
since the interval $[r_0,a]$ can be made
arbitrarily small, the nonzero surface stresses
result in a wormhole that can be said to
approximate a thin-shell wormhole.  (Recall that
a junction surface is called a thin shell
whenever the surface stresses are nonzero,
otherwise a boundary surface.)  This
wormhole differs from the usual cut-and-paste
thin-shell wormhole in the sense that there
exists a complete description in the form of
a line element:
\begin{equation}\label{E:final}
 ds^2=
 \begin{cases}
   -e^{2\Phi(r)}dt^2+\frac{dr^2}{1-b(r)/r}
   +r^2( d\theta^2+\text{sin}^2\theta\, d\phi^2),\\
      \qquad\qquad\qquad\qquad\qquad\qquad\qquad\qquad
      r_0\le r\le a,\\
   -\left(1-\frac{2M}{r}\right)dt^2+\frac{dr^2}{1-2M/r}
   +r^2(d\theta^2+\text{sin}^2\theta\, d\phi^2),\\
       \qquad\qquad\qquad\qquad\qquad\qquad\qquad\qquad
       \qquad
       r>a,
       \end{cases}
\end{equation}
where $b(r)$ is given in Eq. (\ref{E:shape2}) and $M$
in Eq. (\ref{E:mass}), while $\Phi(a)=
\frac{1}{2}\text{ln}(1-2M/a)$.

\section{Stability Analysis}\noindent
It was noted earlier that the junction surface
$r=a$ is a function of proper time $\tau$.
Accordingly, following Lobo and Crawford
\cite{LC05}, the density takes on the form
\begin{equation}\label{E:sigtau}
   \sigma =-\frac{1}{4\pi a}\left(
   \sqrt{1-\frac{2M}{a}+\dot{a}^2}-
   \sqrt{1-\frac{b(a)}{a}+\dot{a}^2}\right),
\end{equation}
where $\dot{a}=da/d\tau$.  A similar adjustment
can be made for $\mathcal{P}$.

To obtain the stability criterion, one starts
by rearranging Eq. (\ref{E:sigtau}),
\[
   \sqrt{1-\frac{2M}{a}+\dot{a}^2}=
   \sqrt{1-\frac{b(a)}{a}+\dot{a}^2}
   -4\pi a\sigma,
\]
in order to obtain the equation of motion
\begin{equation}
   \dot{a}^2+V(a)=0,
\end{equation}
where $V(a)$ is the potential.  It can be shown
that
\begin{equation}\label{E:V}
   V(a)=1-\frac{M+b(a)/2}{a}-\frac{m_s^2}{4a^2}-
   \frac{(M-b(a)/2)^2}{m_s^2},
\end{equation}
using the notation in Ref. \cite{LC05}.  The
idea is to linearize around a static solution
$a=a_0$:
\begin{equation*}
   V(a)=V(a_0)+V'(a_0)(a-a_0)+\frac{1}{2}
   V''(a_0)(a-a_0)^2\\+\text{higher-order terms}.
\end{equation*}
According to Ref. \cite{LC05}, a long and tedious
calculation shows that $V(a_0)=0$ and
$V'(a_0)=0$.  So the stability criterion is
\begin{equation}
     V''(a_0)>0.
\end{equation}

The question now is how to make best use of
Eq. (\ref{E:V}).  Substituting the expression
for $m_s$ from Eq. (\ref{E:ms}) at $a=a_0$
merely confirms that $V(a_0)=0$.  So we are
going to assume that $m_s$ is a constant
and then take advantage of the rather simple
form of the shape function.

A straightforward calculation shows that
\begin{equation}
   V''(a_0)=\\-\frac{2M}{a_0^3}
   -\frac{r_0}{a_0^3}
   -\frac{1}{a_0^3}\left(\frac{8\pi a_0^3}{3}
   -\frac{8\pi r_0^3}{3}\right)\rho_0
   -\frac{3m_s^2}{2a_0^4}\\
   +\frac{1}{m_s^2}\left[(M-\frac{1}{2}
   b(a_0))b''(a_0)-\frac{1}{2}
   (b'(a_0))^2\right].
\end{equation}
In the last term, observe that
$|M-\frac{1}{2}b(a_0)|$ is typically much smaller
than $(b'(a_0))^2$, making all the terms negative.
We conclude that the wormhole is unstable to
linearized radial perturbations, regardless of
how one interprets the parameter $\beta$ in
Sec. \ref{S:Introduction}.

\section{Conclusion}\noindent
A thin-shell wormhole is theoretically
constructible by the so-called cut-and-paste
technique: surgically graft two Schwarzschild
black holes together to form a structure that
is geodesically complete.  The type of
wormhole discussed in this note can be described
by the line element in Eq. (\ref{E:final}).  It
consists of a spherical shell that can be made
arbitrarily thin, and so, by continuity, can
be viewed as an approximation of the cut-and-paste
thin-shell wormhole.  Both are
unstable to linearized radial perturbations.

Because of the instability, it is important to
note that an earlier paper by the author
\cite{pK12} has shown that a
noncommutative-geometry background produces a
small region of stability around $a=a_0$ for a
thin-shell wormhole.  Two earlier papers by
Usmani et al. \cite{aU10} and Rahaman et al.
\cite{fR10} have also addressed the stability
issue.  According to Ref. \cite{aU10}, a
stable region exists for thin-shell wormholes
constructed from black holes in generalized
dilaton-axion gravity.  Analogous results
hold for regular charged black holes
\cite{fR10}.  (The stability region is
shown correctly in Ref. \cite{fR10} but
not in the arXiv version.)  So under
certain conditions, thin-shell wormholes
can be stable to linearized radial
perturbations.

\pagebreak

\end{document}